\documentclass[aps,pra,eqsecnum,amsmath,amsfonts,12pt,tightenlines]{revtex4}
\usepackage{amssymb}
\usepackage{graphicx}

\newcommand{\ket}[1]{|#1\rangle}
\newcommand{\bra}[1]{\langle{#1}|}

\newcommand{\ketbra}[1]{\ket{#1}\bra{#1}}

\newcommand{\abs}[1]{\left|#1\right|}
\newcommand{\mc}[1]{\mathcal{#1}}

\def\co{\mathord{\mbox{co}}}
\def\vec#1{\mathbf{#1}}

\begin{document}

\title{Entanglement of Formation of Rotationally Symmetric States}
\author{Kiran K.~Manne and Carlton M.~Caves\vspace{0.1in}}
\affiliation{Department of Physics and Astronomy, University of New Mexico,\\
Albuquerque, New Mexico 87131--1156, USA\vspace{0.1in}}

\date{2005 June~17}

\begin{abstract}
Computing the entanglement of formation of a bipartite state is
generally difficult, but special symmetries of a state can
simplify the problem. For instance, this allows one to determine
the entanglement of formation of Werner states and isotropic
states. We consider a slightly more general class of states,
rotationally symmetric states, also known as SU(2)-invariant
states. These states are invariant under global rotations of both
subsystems, and one can examine entanglement in cases where the
subsystems have different dimensions.  We derive an analytic
expression for the entanglement of formation of rotationally
symmetric states of a spin-$j$ particle and a spin-${1\over2}$
particle. We also give expressions for the I-concurrence,
I-tangle, and convex-roof-extended negativity.
\end{abstract}

\maketitle

\section{Introduction}
\label{section:intro} An entangled pure state is one in which
complete knowledge about the overall state is incomplete with
regard to the subsystems~\cite{schroedinger}.  Such states are
strongly correlated in the sense that the correlations cannot be
reproduced by a local hidden variable theory~\cite{epr,bell}.
Besides being integral ingredients in no-go theorems regarding the
interpretation of quantum mechanics, entangled states open new
avenues in quantum engineering and information
processing~\cite{lo,nielsen}. Using entangled states, we can
teleport quantum information, we can in principle build quantum
computers that provide exponential speedup over classical
computers, and we can also prove the security of assorted
cryptographic protocols~\cite{acin03}. We can do all this and more
without a complete understanding of entanglement, because the
tasks rely only on pure states.  Yet one often must deal with
mixed states that arise from physical processes such as noise.
This makes it important to have a measure of entanglement for
mixed states so that potential applications can be explored and
evaluated.

A commonly used entanglement measure for a pure state $\ket\psi$ of two
systems, $A$ and $B$, is the entropy of the marginal density operator
$\rho_A$ (or $\rho_B$),
\begin{equation}
\label{eq:puremeasure}
E(\psi) = S(\rho_A)= -\mbox{tr}(\rho_A\log\rho_A)
=-\mbox{tr}(\rho_B\log\rho_B)\;.
\end{equation}
The importance of this measure comes chiefly from the fact that it
gives the rate at which copies of a pure state can be converted,
by using only local operations and classical communication
(LOCCs), into copies of maximally entangled states and vice
versa~\cite{bennett96-1,popescu97}. The entanglement of
formation~\cite{bennett96-1,bennett96-2},
\begin{equation}
\label{eq:eof}
E_F(\rho) \equiv
\min_{\{p_j,\ket{\psi_j}\}}
\Biggl(\sum_j p_jE(\psi_j)\Biggm|\rho=\sum_jp_j\ketbra{\psi_j}\Biggr)\;,
\end{equation}
is the so-called convex-roof extension of the pure-state measure
(\ref{eq:puremeasure}) to bipartite mixed states $\rho$.  The need
to search over all ensemble decompositions of $\rho$ generally
makes it impossible to calculate the entanglement of formation,
unless some efficient method to short-circuit a complete search
can be found.  The entanglement of formation provides an upper
bound on the rate at which the maximally entangled states must be
supplied to create copies of $\rho$~\cite{wootters01, hayden01}
and the rate at which one can distill maximally entangled states
from $\rho$. It is also a measure for which analytic results are
known. The entanglement of formation is known for arbitrary
bipartite qubit states~\cite{wootters98}, isotropic
states~\cite{terhal00} and Werner states~\cite{vollbrecht02} in
arbitrary dimensions, and symmetric gaussian states in infinite
dimensions~\cite{giedke03}.

Many other mixed-state entanglement measures have been defined:
entanglement of distillation~\cite{bennett96-1,bennett96-2},
negativity~\cite{vidal02}, relative entropy~\cite{vedral96}, robustness
of entanglement~\cite{vidal99}, I-concurrence~\cite{rungta01},
I-tangle~\cite{osborne02}, the geometric measure~\cite{wei03}, and
others. Each is useful in particular physical contexts, but the
different measures produce different orderings on mixed states even
when they agree on pure states~\cite{virmani00}. Except for negativity,
these measures are also difficult to calculate for an arbitrary mixed
state.  Here we focus on the entanglement of formation, but provide
results for other measures since our results on entanglement of
formation can easily be extended to these other measures.

Terhal and Vollbrecht~\cite{terhal00} showed how to use the symmetry of
the isotropic states to find the entanglement of those states.
Vollbrecht and Werner~\cite{vollbrecht02} elucidated and extended the
results of that paper; in particular, they formulated a general
technique that leverages symmetries to simplify the calculation of
convex-roof measures for symmetric states.  We apply their method to a
bipartite system consisting of a spin-$j$ particle and a
spin-${1\over2}$ particle, i.e., a $[(2j+1)\times2]$-dimensional
Hilbert space, where the states we consider are those that are
invariant under global rotations of the two particles.  Such states are
also known as SU(2)-invariant states~\cite{schliemann02}. They are
functions of a single parameter $p$,
\begin{equation}
\label{eq:invariantstate}
\rho(p)=\frac{1-p}{2j+2}\Pi_{j+1/2}+\frac{p}{2j}\Pi_{j-1/2}\;,
\end{equation}
where the operators
\begin{equation}
\label{eq:angmomproj}
\Pi_{j\pm1/2}=\sum_{m=-(j\pm1/2)}^{j\pm1/2}\ketbra{j\pm\textstyle{{1\over2}},m},
\end{equation}
are the projectors onto the subspaces of total angular momentum
$j\pm{1\over2}$.  These states can appear when one loses
information regarding the Cartesian reference frame of the two
systems~\cite{bartlett03}.  They can also arise from bipartite
splits of a rotationally symmetric chain of spin-$1\over2$
particles into a single qubit and the rest.  Such states can also
arise as the multiphoton states that are generated by parametric
down-conversion and then undergo photon losses~\cite{durkin04}.

The rotationally symmetric state~(\ref{eq:invariantstate}) is known to
be separable if and only if~\cite{schliemann02,hendriks}
\begin{equation}
p\leq\frac{2j}{2j+1}\;.
\end{equation}
In this paper we show that the entanglement of formation of the
SU(2)-invariant state~(\ref{eq:invariantstate}) is given by
\begin{equation}
\label{eq:EoF} E_F\bigl(\rho(p)\bigr) =
\begin{cases}
0\;,&p\in[0,2j/(2j+1)],\\
H\!\left(\displaystyle{\frac{1}{2j+1}}\Bigl(\sqrt{p}-\sqrt{2j(1-p)}\,\Bigr)^2\right)\;,&
p\in[2j/(2j+1,1].
\end{cases}
\end{equation}
Here $H(x)=-x\log x-(1-x)\log(1-x)$ is the binary entropy; we use the
natural logarithm in calculating actual values of the entanglement of
formation and when taking derivatives of entropic expressions.  The
formula~(\ref{eq:EoF}) is noteworthy in that it provides the first
example of an entanglement of formation for subsystems having different
dimensions.

In Sec.~\ref{section:symmetry}, we briefly review the
Terhal-Vollbrecht-Werner program for determining the entanglement of
symmetric states. Section~\ref{section:su2} considers rotationally
invariant states and their properties.  In Sec.~\ref{section:eof}, we
calculate the entanglement of formation of rotationally invariant
states of a spin-$j$ particle and a spin-${1\over2}$ particle.
Section~\ref{section:others} contains analytic expressions for other
mixed-state entanglement measures. Finally, in
Sec.~\ref{section:summary} we address possible extensions of our
results.

\section{Convex-Roof Entanglement Measures Under Symmetry}
\label{section:symmetry}

In this section we summarize the Terhal-Vollbrecht-Werner
procedure~\cite{vollbrecht02} for determining the entanglement of
symmetric states.

We begin with a few definitions.  Let $K$ be a compact convex set
(e.g., density matrices), which is itself a subset of a
finite-dimensional vector space $V$, and let $M\subset K$ be an
arbitrary subset of $K$ (e.g.,~pure states).  Let $f:M\rightarrow
\mathbb{R}\cup\{+\infty\}$ be a real-valued function on $M$ (e.g., an
entanglement measure on pure states).  The convex roof of $f$, $\co
f:K\rightarrow\mathbb{R}\cup\{+\infty\}$, is a function on the entire
set $K$, defined by
\begin{equation}
\co f(x)\equiv
\min_{\{s_j\in M\}}
\Biggl(
\sum_j\lambda_j f(s_j)
\Biggm|
\sum_j\lambda_j s_j = x,\;
\lambda_j\geq 0,\;\sum_j\lambda_j =1
\Biggr)\;.
\label{eq:conroof}
\end{equation}
The convex roof is computed by minimizing the average value of $f$ over
all possible ways of writing an element $x\in K$ as a convex
combination of elements in~$M$. In the case of density operators and
pure states, such a convex combination is called a pure-state ensemble
decomposition, so we refer generally to a decomposition of $x$.  If $x$
has no decomposition in terms of elements of $M$, $\co f(x)$ is
infinite.  The convex roof $\co f(x)$ is the largest convex function
$g$ on $K$ such that $g(s)\le f(s)$ for $s\in M$; if $M$ is a set of
extreme points of $K$, as in the case of pure states and density
operators, $\co f(s)=f(s)$ for $s\in M$.  In this notation, the
entanglement of formation~(\ref{eq:eof}) is
\begin{equation}
E_F(\rho) = \co\,E(\rho)\;.
\end{equation}

Now suppose there exists a symmetry group $\mc{G}$ that acts on $V$
through its matrix representations $\alpha_G$.  Assume that $\alpha_G
K\subset K$ and also that $\alpha_G M\subset M$, with
$f\bigl(\alpha_G(s)\bigr) = f(s)$ for all $s\in M$.  In the case of
entanglement, the linear transformations $\alpha_G$ are tensor products
of local unitary representations of a symmetry group; as unitary
transformations, they take pure states to pure states, and as tensor
products of local unitary transformations on the two subsystems, they
preserve the entanglement measure.  Define a projection
$\mathbf{P}:K\rightarrow K$ by averaging uniformly over the group,
\begin{equation}
\mathbf{P}(x)\equiv\int \mbox{d}G\,\alpha_G(x)\;.
\end{equation}
The projection $\mathbf{P}$ is often called \emph{twirling}.  We denote
the range of $\mathbf{P}$ by $\mathbf{P}K$.  Elements of $\mathbf{P}K$
are precisely those elements of $K$ that are invariant under the
projection.  More generally, $\mathbf{P}K$ consists of all elements of
$K$ that are invariant under the group operations and thus are known as
\emph{group-invariant elements}.  It is clear that $\mathbf{P}K$ is a
convex set.

The problem of finding $\co f(x)$ for $x\in\mathbf{P}K$ is simplified,
because the problem can be divided into two parts.  One first
determines the function $\epsilon:\mathbf{P}K\rightarrow
\mathbb{R}\cup\{+\infty\}$ defined by
\begin{equation}
\label{eq:epsilon}
\epsilon(x)\equiv\min_{\{s\in M\}}\bigl(f(s)\bigm|\mathbf{P}(s) = x\bigr)\;.
\end{equation}
Thus instead of minimizing over all possible decompositions, one first
minimizes over a subset of the pure states.  Then the convex roof of
$f$ on $\mathbf{P}K$ is the convex hull of $\epsilon\,$:
\begin{equation}
\co f(x) = \co\,\epsilon(x)\;,\quad x\in\mathbf{P}K\;.
\label{eq:cofepsilon}
\end{equation}
The convex hull of $\epsilon$ is the largest convex function on
$\mathbf{P}K$ that nowhere exceeds $\epsilon$.  We use the same
notation for convex hull and convex roof, because the convex hull of a
function $f$ is the convex roof for the special case $M=K$, i.e., in
this case,
\begin{equation}
\co\,\epsilon(x)=
\min_{\{x_j\in\mathbf{P}K\}}
\Biggl(
\sum_j\lambda_j\epsilon(x_j)
\Biggm|
\sum_j\lambda_j x_j = x,\;
\lambda_j\geq 0,\;\sum_j\lambda_j =1
\Biggr)\;.
\label{eq:conroofepsilon}
\end{equation}
For highly symmetric states,
the more difficult of the two steps is determining $\epsilon(x)$, since
finding $\co\,\epsilon(x)$ is usually straightforward or even
unnecessary because $\epsilon(x)$ is already convex.

Demonstrating the reduction of the minimization~(\ref{eq:conroof}) to
Eqs.~(\ref{eq:epsilon}) and (\ref{eq:cofepsilon}) is sufficiently
simple that we include it here for completeness.  First, let
$x\in\mathbf{P}K$ have the optimal decomposition $x=\sum_j\lambda_j
s_j$ relative to Eq.~(\ref{eq:conroof}), i.e., $\co
f(x)=\sum_j\lambda_j f(s_j)$.  Defining $x_j=\mathbf{P}(s_j)$, we have
$\epsilon(x_j)\le f(s_j)$, $x=\mathbf{P}(x)=\sum_j\lambda_j x_j$, and
\begin{equation}
\co f(x)=
\sum_j\lambda_j f(s_j)\ge\sum_j\lambda_j\epsilon(x_j)
\ge\co\,\epsilon(x)\;.
\label{eq:first}
\end{equation}
Second, let $x\in\mathbf{P}K$ have the optimal decomposition
$x=\sum_j\lambda_j x_j$ relative to Eq.~(\ref{eq:conroofepsilon}),
i.e., $\co\,\epsilon(x)=\sum_j\lambda_j\epsilon(x_j)$, and let
$s_j$ achieve the minimum in Eq.~(\ref{eq:epsilon}) for $x_j$,
i.e., $x_j=\mathbf{P}(s_j)$ and $\epsilon(x_j)=f(s_j)$.  Then
\begin{equation}
x=\sum_j\lambda_j\int\mbox{d}G\,\alpha_G(s_j)
\end{equation}
is a decomposition of $x$, and
\begin{equation}
\co f(x)\le
\sum_j\lambda_j\int\mbox{d}G\,f\bigl(\alpha_G(s_j)\bigr)=
\sum_j\lambda_j f(s_j)=
\sum_j\lambda_j\epsilon(x_j)=
\co\,\epsilon(x)\;.
\label{eq:second}
\end{equation}
Together, Eqs.~(\ref{eq:first}) and (\ref{eq:second}) give
Eq.~(\ref{eq:cofepsilon}).

This method has been used to compute the entanglement of the
Werner states~\cite{vollbrecht02}, which are invariant under
$U\otimes U$, and the isotropic states~\cite{terhal00}, which are
invariant under $U\otimes U^*$. One can generalize to a larger
class of states by considering a subgroup of either $U\otimes U$
or $U\otimes U^*$. In the next section we examine groups of the
form $R\otimes R$, where $R$ is a rotation.

\section{SU(2)-Invariant States}
\label{section:su2}

Consider two particles, one of spin~$j_1$ and the other of spin $j_2$.
We consider states that are symmetric under global rotations.  The
symmetry group is $\mc{R} = \{D^{(j_1)}(R)\otimes D^{(j_2)}(R)\}$,
where $D^{(j)}(R)=\exp(-i\theta \vec J\cdot\vec n)$ is the spin-$j$
representation of the rotation $R\in SO(3)$ [or equivalently in
$SU(2)$].  Here $\vec J$ is the angular-momentum vector, with
components $J_x$, $J_y$, and $J_z$.  The twirling operator is then
\begin{equation}
\mathbf{P}_\mc{R}(\rho)=
\int \mbox{d}\mu(R)\,
D^{(j_1)}(R)\otimes D^{(j_2)}(R)\rho D^{(j_1)}(R)^\dagger
\otimes D^{(j_2)}(R)^\dagger\;,
\end{equation}
where $\mu(R)$ is the group-invariant measure for the rotation group.
The twirling operation describes a process where two parties use
classical communication to select a random rotation $R$ that each party
implements locally.  This makes twirling a LOCC operation, which
implies that entanglement does not increase under twirling.

The states that are invariant under the twirling operation are those
that are convex combinations of the states associated with the
projectors onto the irreducible subspaces of total angular momentum $J
=\abs{j_1-j_2},\abs{j_1-j_2}+1,\ldots,j_1+j_2$.  These projectors,
\begin{equation}
\Pi_J =\sum_{m=-J}^{J}\ketbra{J,m}\;,
\end{equation}
have associated normalized states $\Pi_J/(2J+1)$.  Thus the
$\mc{R}$-invariant states are
\begin{equation}
\label{eq:generalinvariant}
\rho(\vec{p}) =
\sum_{J=\abs{j_1-j_2}}^{j_1+j_2}\frac{p_J}{2J+1}\Pi_J\;,
\end{equation}
where $p_J=\mbox{tr}\bigl(\rho(\vec p)\Pi_J\bigr)\geq 0$ and $\sum
p_J =1$. Any state $\sigma$ twirls to a $\mc{R}$-invariant state,
i.e., $\mathbf{P}_{\mc{R}}(\sigma)=\rho(\vec p)$, where
$p_J=\mbox{tr}\bigl(\mathbf{P}_{\mc{R}}(\sigma)\Pi_J\bigr)=
\mbox{tr}\bigl(\sigma\mathbf{P}_{\mc{R}}(\Pi_J)\bigr)$.  Since
$\Pi_J$ is $\mc{R}$ invariant, we find that
$p_J=\mbox{tr}(\sigma\Pi_J)$ is the overlap of $\sigma$ with the
subspace of total angular momentum~$J$.  Notice also that
\begin{equation}
\mathbf{P}_{\mc R}\bigl(\ketbra{J,m}\bigr)=
{\Pi_J\over2J+1}
\end{equation}
for any value of $m$.

We note that, for the $\mc{R}$-invariant
state~(\ref{eq:generalinvariant}), the positive partial transpose
condition is necessary and sufficient for separability when
$j_1={1\over2}$ and $j_2$ is arbitrary and when $j_1=1$ and $j_2$ is an
integer~\cite{schliemann02,hendriks,schliemann05,breuer05a,breuer05b}.

The general problem of finding the entanglement of formation of
rotationally invariant states is first to determine the function
\begin{equation}
\epsilon(\vec p)=
\min_{\{|\psi\rangle\}}
\Bigl(E(\psi)\Bigm|\mathbf{P}_{\mc{R}}(|\psi\rangle\langle\psi|)=\rho(\vec p)\Bigr)
=\min_{\{|\psi\rangle\}}
\bigl(\,E(\psi)\bigm|\langle\psi|\Pi_j|\psi\rangle=p_j\,\bigr)
\end{equation}
and then to find its convex hull on the convex set of probabilities
$\vec p$.

It is interesting to note that although states invariant under
$U\otimes U$ and $U\otimes U^*$ are quite different, this is not the
case if one compares states invariant under $R\otimes R$ and $R\otimes
R^*$. The reason is that conjugation of a representation is equivalent
to rotating by $\pi$ about the $y$ axis, i.e., $D^{(j)}(R)^* = e^{i\pi
J_y}D^{(j)}(R) e^{-i\pi J_y}$.  Thus the states that are invariant
under $\widetilde{\mc{R}} = \{D^{(j_1)}(R)\otimes D^{(j_2)}(R)^*\}$ are
obtained from the rotationally invariant states by rotating one of the
two systems by $\pi$ about the $y$ axis.

\section{Entanglement of Formation for SU(2)-Invariant States of Spin-$j$
and Spin-$1\over2$ Particles}
\label{section:eof}

\subsection{General considerations}

To obtain insight into the entanglement of rotationally symmetric
states, we examine the simplest case, a particle of arbitrary spin
$j$ and a particle of spin-${1\over2}$.  The eigenstates of total
angular momentum are well known, given by the Clebsch-Gordon
coefficients~\cite{shankar}, and can be written as
\begin{equation}
\ket{j\pm\textstyle{\frac{1}{2}},m}=
\pm\sqrt{\displaystyle{j+{1\over2}\pm m\over2j+1}}
\,\ket{j,m-\textstyle{\frac{1}{2}}}\otimes
\ket{\textstyle{\frac{1}{2},\frac{1}{2}}}
+\sqrt{\displaystyle{j+{1\over2}\mp m\over2j+1}}
\,\ket{j,m+\textstyle{\frac{1}{2}}}\otimes
\ket{\textstyle{\frac{1}{2}},-\textstyle{\frac{1}{2}}}\;.
\end{equation}
It is straightforward to calculate the
entanglement~(\ref{eq:puremeasure}) of these states:
\begin{equation}
E\bigl(\ket{j\pm\textstyle{1\over2},m}\bigr)
=H\!\left(\displaystyle{{j+{1\over2}\pm m\over2j+1}}\right)
=H\!\left(\displaystyle{\frac{1}{2}-\frac{|m|}{2j+1}}\right)\;.
\end{equation}
Eigenstates with identical values of $\abs{m}$ have the same
entanglement.  In the $j+{1\over2}$ subspace, the minimum entanglement
is $0$, achieved when $|m|=j+{1\over2}$ and in the $j-{1\over2}$
subspace, the minimum entanglement is $H\bigl(1/(2j+1)\bigr)$, achieved
when $|m|=j-{1\over2}$.

The $\mc{R}$-invariant states are convex combinations of the states
associated with the projectors~(\ref{eq:angmomproj}) onto the subspaces
of total angular momentum $j+{1\over2}$ and $j-{1\over2}$,
\begin{equation}
\rho(p) = \frac{1-p}{2(j+1)}\Pi_{j+1/2} + \frac{p}{2j}\Pi_{j-1/2}\;,
\end{equation}
where $p=\mbox{tr}\bigl(\rho(p)\Pi_{j-1/2}\bigr)$.  Any state $\sigma$
twirls to $\mathbf{P}_{\mc{R}}(\sigma)=\rho(p)$, where
$p=\mbox{tr}(\sigma\Pi_{j-1/2})$ is the overlap of $\sigma$ with the
$j-{1\over2}$ subspace.

The first reason this problem is simpler than the general case of two
arbitrary spins is that the rotationally invariant states are specified
by the single parameter $p$.  The problem of finding the entanglement
of formation reduces to determining a function of this one parameter,
\begin{equation}
\epsilon(p)
=\min_{\{|\psi\rangle\}}
\bigl(\,E(\psi)\bigm|\langle\psi|\Pi_{j-1/2}|\psi\rangle=p\,\bigr)\;,
\label{eq:epsilonp}
\end{equation}
and then finding its convex hull, $\co\,\epsilon(p)$.

\subsection{Determining $\epsilon(p)$}
To find $\epsilon(p)$, we begin by looking at a couple of example
states.  The first,
\begin{eqnarray}
\ket\phi
&=&\ket{j,j}\otimes
\bigl(\sqrt{1-\nu}\,\ket{\textstyle{{1\over2}},\textstyle{{1\over2}}}
+\sqrt\nu\,\ket{\textstyle{{1\over2}},\textstyle{-{1\over2}}}
\bigr)\nonumber\\
&=&\sqrt{1-\nu}\,\ket{j+\textstyle{{1\over2}},j+\textstyle{{1\over2}}}
+\sqrt{\displaystyle{{\nu\over2j+1}}}\ket{j+\textstyle{{1\over2}},j-\textstyle{{1\over2}}}
+\sqrt{\displaystyle{{2j\nu\over2j+1}}}\ket{j-\textstyle{{1\over2}},j-\textstyle{{1\over2}}}\;,
\end{eqnarray}
is a product state, thus having no entanglement, i.e.,
$E(\phi)=0$. This state has overlap
$p=\langle\phi|\Pi_{j-1/2}|\phi\rangle=2j\nu/(2j+1)$.  As $\nu$
ranges from 0 to 1, $p$ varies from 0 to $2j/(2j+1)$, which shows
that $\epsilon(p)=0$ for $p\in[0,2j/(2j+1)]$. Thus in determining
$\epsilon(p)$, we can restrict our attention to
$p\in[2j/(2j+1),1]$.

The second state,
\begin{eqnarray}
\ket{\chi}&=&
-\sqrt{\mu}\,\ket{j,j-1}\otimes\textstyle{\ket{\frac{1}{2},\frac{1}{2}}}+
\sqrt{1-\mu}\,\ket{j,j}\otimes\textstyle{\ket{\frac{1}{2},-\frac{1}{2}}}\nonumber\\
&=&{1\over\sqrt{2j+1}}
\Bigl[
\bigl(-\sqrt{2j\mu}+\sqrt{1-\mu}\,\bigr)\ket{j+\textstyle{{1\over2}},j-\textstyle{{1\over2}}}
+\bigl(\sqrt{\mu}+\sqrt{2j(1-\mu)}\,\bigr)\ket{j-\textstyle{{1\over2}},j-\textstyle{{1\over2}}}
\Bigr]\;,\nonumber\\
\label{eq:minstate}
\end{eqnarray}
is entangled, with $E(\chi)=H(\mu)$, and has overlap
\begin{equation}
p=\langle\chi|\Pi_{j-1/2}|\chi\rangle=
\frac{1}{2j+1}\left(\sqrt\mu+\sqrt{2j(1-\mu)}\,\right)^2\equiv p_\mu\;.
\label{eq:pmu}
\end{equation}
As $\mu$ increases from 0, $p$ increases monotonically from a value of
$2j/(2j+1)$ at $\mu=0$ to a maximum value of $1$ at $\mu=1/(2j+1)$; for
larger values of $\mu$, $p$ decreases monotonically to a value of
$1/(2j+1)$ at $\mu=1$.

Inverting to find $\mu$ as a function of $p$ and using the branch that
gives the smaller values of $\mu$, we find that for
$p\in[2j/(2j+1),1]$,
\begin{equation}
\mu=\frac{1}{2j+1}\left(\sqrt p-\sqrt{2j(1-p)}\,\right)^2
\equiv\mu_{\rm min}(p)\;.
\label{eq:muinvert}
\end{equation}
The reason for the functional notation $\mu_{\rm min}(p)$ becomes clear
below.  The function~$\mu_{\rm min}(p)$ increases monotonically from a
value of $\mu=0$ at $p=2j/(2j+1)$ to $\mu=1/(2j+1)\le1/2$ at $p=1$. The
upshot is that the state~(\ref{eq:minstate}) tells us that for
$p\in[2j/(2j+1),1]$,
\begin{equation}
\label{eq:epsilonbound}
\epsilon(p)\le H\bigl(\mu_{\rm min}(p)\bigr)=
H\!\left(\displaystyle{\frac{1}{2j+1}}\Bigl(\sqrt{p}-\sqrt{2j(1-p)}\,\Bigr)^2\right)
\le H\!\left({1\over2j+1}\right)\;.
\end{equation}
It turns out that $\ket\chi$ achieves the minimum value of
$E(\psi)$ in Eq.~(\ref{eq:epsilonp}) and thus that $\epsilon(p)$
is actually given by the expression in the middle of
Eq.~(\ref{eq:epsilonbound}). To proceed with the proof of this,
however, the only information we need from $\ket\chi$ is the final
inequality in Eq.~(\ref{eq:epsilonbound}), i.e., $\epsilon(p)\le
H\bigl(1/(2j+1)\bigr)$.

The next step is to characterize the set of pure states over which we
must minimize in Eq.~(\ref{eq:epsilonp}).  Any pure state of a
$(2j+1)\times2$ system can be written in terms of a two-term Schmidt
decomposition,
\begin{equation}
\ket{\psi}=
\sqrt{\mu}\,\ket{e_1}\otimes\ket{f_1} +
\sqrt{1-\mu}\,\ket{e_2}\otimes\ket{f_2})\;.
\label{eq:Schmidt}
\end{equation}
which has entanglement $E(\psi)=H(\mu)$.  The second reason this
problem is simpler than the general case is that the pure states that
twirl to invariant states are specified by the single parameter $\mu$.
The final inequality in Eq.~(\ref{eq:epsilonbound}) implies that the
only values of $\mu$ we need to consider are $\mu\in(0,1/(2j+1)]$, for
which $H(\mu)$ is monotonically increasing.

We can introduce two unitary operators, $V$ on the spin-$j$ particle
and $W$ on the spin-$1\over2$ particle, which transform bases of our
choice to the Schmidt bases. In particular, we can write $\ket\psi$ as
\begin{equation}
\ket\psi=
V\otimes W\ket\chi=
V\otimes W
\bigl(-\sqrt{\mu}\,\ket{j,j-1}\otimes\textstyle{\ket{\frac{1}{2},\frac{1}{2}}}+
\sqrt{1-\mu}\,\ket{j,j}\otimes\textstyle{\ket{\frac{1}{2},-\frac{1}{2}}}\bigr)\;.
\end{equation}
Any unitary on the spin-$1\over2$ system is equivalent (up to an
irrelevant phase) to a rotation $R$, i.e., $W=D^{(1/2)}(R)$.  Rotating
both particles by the inverse of $R$, we obtain the state
\begin{equation}
\label{eq:fiducial}
\ket{\psi_\mu(U)}=
U\otimes I\ket\chi=
U\otimes I
\bigl(-\sqrt{\mu}\,\ket{j,j-1}\otimes\textstyle{\ket{\frac{1}{2},\frac{1}{2}}}+
\sqrt{1-\mu}\,\ket{j,j}\otimes\textstyle{\ket{\frac{1}{2},-\frac{1}{2}}}\bigr)\;,
\end{equation}
where $U=D^{(j)}(R)^\dagger V$.

The significance of this move is that $\ket{\psi_\mu(U)}$ has the same
entanglement and the same overlap with the $j-\textstyle{1\over2}$
subspace as does $\ket\psi$.  Thus in doing the
minimization~(\ref{eq:epsilonp}), we only need to consider the
states~$\ket{\psi_\mu(U)}$, i.e.,
\begin{equation}
\epsilon(p)
=\min_{\{\mu,U\}}
\bigl(\,H(\mu)\bigm|p_\mu(U)=p\,\bigr)\;,
\label{eq:epsilonp2}
\end{equation}
where
\begin{equation}
p_\mu(U)=\langle\psi_\mu(U)|\Pi_{j-1/2}|\psi_\mu(U)\rangle
=\langle\chi|U^\dagger\otimes I\,\Pi_{j-1/2}U\otimes I|\chi\rangle\;.
\end{equation}
Moreover, since $H(\mu)$ is a monotonically increasing function of
$\mu$ in the interval of interest, we can simply minimize $\mu$ over
all unitaries $U$,
\begin{equation}
\tilde{\mu}(p)\equiv\min_{\{U\}} \bigl(\,\mu\bigm|p_\mu(U)=p\,\bigr)\;,
\label{eq:mumin}
\end{equation}
and plug the result into the binary entropy to give
\begin{equation}
\epsilon(p)=H\bigl(\tilde{\mu}\bigr)\;. \label{eq:Hmumin}
\end{equation}
The final reason that the problem of finding the entanglement of
formation in this case is doable is that the problem can be reduced to
the minimization~(\ref{eq:mumin}) over a single unitary, that being a
unitary acting on the spin-$j$ particle.

Our strategy now is to show that $p_\mu(U)\le p_\mu(I)=p_\mu$, where
$p_\mu$ is the function~(\ref{eq:pmu}). This result, which we will
prove below, allows us to immediately determine $\tilde{\mu}(p)$.  Since
$p_\mu$ is monotonically increasing on the relevant interval
$(0,1/(2j+1)]$, we can find a $\mu'\le\mu$ such that
$p_\mu(U)=p_{\mu'}(I)$ for any $\mu$ and $U$, implying that the minimum
is always achieved by $U=I$. This means that $\tilde{\mu}$ is obtained
by inverting $p_{\tilde{\mu}}=p$, and thus
\begin{equation}
\tilde{\mu}(p)=\mu_{\rm min}(p)\;,
\end{equation}
as given by Eq.~(\ref{eq:muinvert}).  As promised, $\epsilon(p)$
on the interval $[2j/(2j+1),1]$ is given by the middle expression
in Eq.~(\ref{eq:epsilonbound}).  The result is that
\begin{equation}
\label{eq:epsilonresult} \epsilon(p)=
\begin{cases}
0\;,&p\in[0,2j/(2j+1)],\\
H\!\left(\displaystyle{\frac{1}{2j+1}}\Bigl(\sqrt{p}-\sqrt{2j(1-p)}\,\Bigr)^2\right)\;,&
p\in[2j/(2j+1),1].
\end{cases}
\end{equation}

We now must show that $p_\mu(U)\le p_\mu$.  We begin by noting that
\begin{eqnarray}
p_\mu(U)&=&
\sum_{m=-j+1/2}^{j-1/2}
\bigl|\langle\chi|U^\dagger\otimes I|j-\textstyle{{1\over2}},m\rangle\bigr|^2\nonumber\\
&=&\displaystyle{{1\over2j+1}
\sum_{m=-j+1/2}^{j-1/2}
\left|
\sqrt{(1-\mu)r_m}\,x_me^{i\alpha_m}
+\sqrt{\mu r_{-m}}\,y_me^{i\beta_m}
\right|^2}\;,
\label{eq:overlap}
\end{eqnarray}
where we define two rows of matrix elements of $U^\dagger$,
\begin{eqnarray}
\bra{j,j}U^\dagger\ket{j,m+\textstyle{{1\over2}}} & = & x_m e^{i\alpha_m}\;,
\quad x_m\ge0\;,\quad m=-j-\textstyle{1\over2},\ldots,j-\textstyle{1\over2}\;,\\
\bra{j,j-1}U^\dagger\ket{j,m-\textstyle{{1\over2}}}& = & y_m e^{i\beta_m}\;,
\quad y_m\ge0\;,\quad m=-j+\textstyle{1\over2},\ldots,j+\textstyle{1\over2}\;,
\end{eqnarray}
and where
\begin{equation}
r_m\equiv j+\textstyle{1\over2}+m\ge0\;.
\end{equation}
Only two rows of the unitary matrix are involved in the
overlap~(\ref{eq:overlap}) because one particle is a qubit.  Now we can
write
\begin{equation}
p_\mu(U)\leq
\frac{1}{2j+1}
\sum_{m=-j+1/2}^{j-1/2}
\Bigl(
\sqrt{(1-\mu)r_m}\,x_m+\sqrt{\mu r_{-m}}\,y_m
\Bigr)^2
\equiv\frac{1}{2j+1}F(x,y)\;,
\label{eq:Fxy}
\end{equation}
with equality when all the phase factors are equal to 1.  For the
remainder of the proof, we omit the range of the sum since it is always
that of Eq.~(\ref{eq:Fxy}), i.e., $m=-j+{1\over2},\ldots,j-{1\over2}$.
Notice that on this range, $r_m$ and $r_{-m}$ are strictly positive.

The problem now is one of maximizing $F(x,y)$ subject to the
constraints on the two relevant rows of $U^\dagger$.  Normalization of
these rows gives the constraints
\begin{equation}
\sum x_m^2=1-x_{-j-1/2}^2
\quad\mbox{and}\quad
\sum y_m^2=1-y_{j+1/2}^2\;.
\end{equation}
The two rows must also be orthogonal, but we ignore this constraint on
the grounds that doing so can only lead to a bigger maximum, which is
still an upper bound for $p_\mu(U)$.  With this constraint ignored, it
is clear from the form of Eq.~(\ref{eq:Fxy}) that the maximum is
achieved when $x_{-j-1/2}=0= y_{j+1/2}$.  Thus we introduce two
Lagrange multipliers and maximize the function
\begin{equation}
G(x,y) = F(x,y) +
\lambda_1\!\left(\sum x_m^2-1\right)
+\lambda_2\!\left(\sum y_m^2-1\right)\;.
\end{equation}
It turns out that solution to this maximization satisfies the
orthogonality constraint.

Setting the first derivatives equal to zero yields the following
equations,
\begin{eqnarray}
[(1-\mu)r_m+\lambda_1]x_m+\sqrt{(1-\mu)\mu r_mr_{-m}}\,y_m & = & 0\;,\nonumber\\
\sqrt{(1-\mu)\mu r_mr_{-m}}\,x_m+(\mu r_{-m}+\lambda_2)y_m & = & 0\;,
\label{eq:derivatives}
\end{eqnarray}
which hold for all $m$.  Since $x_m$ and $y_m$ are both nonnegative, we
must have $\lambda_1\le-(1-\mu)r_m<0$ and $\lambda_2\le-\mu r_{-m}<0$.
Moreover, if the solution is not to be $x_m=y_m=0$, the determinant of
the matrix of coefficients must vanish, which gives
\begin{equation}
\lambda_1\lambda_2+(1-\mu)r_m\lambda_2+\mu r_{-m}\lambda_1=0\;.
\label{eq:determinant}
\end{equation}

Now suppose Eq.~(\ref{eq:determinant}) holds for two or more values of
$m$.  By considering any pair of $m$ values for which
Eq.~(\ref{eq:determinant}) holds, it is easy to show that the Lagrange
multipliers are given by $\lambda_1=\lambda_2=0$ or by
\begin{eqnarray}
\lambda_1&=&-(1-\mu)(r_m+r_{-m})=-(1-\mu)(2j+1)\;,\label{eq:lambda1}\\
\lambda_2&=&-\mu(r_m+r_{-m})=-\mu(2j+1)\label{eq:lambda2}\;.
\end{eqnarray}
The former case is ruled out by the requirement that $\lambda_1$ and
$\lambda_2$ be negative, so we need only consider Eqs.~(\ref{eq:lambda1})
and (\ref{eq:lambda2}), which imply
\begin{equation}
x_m=\sqrt{{\mu\over1-\mu}}\sqrt{{r_m\over r_{-m}}}\,y_m\;.
\end{equation}
Now we find that
\begin{equation}
\sum x_m^2={\mu\over1-\mu}\sum{r_m\over r_{-m}}y_m^2
\le{2j\mu\over1-\mu}\sum y_m^2\le1\;,
\end{equation}
with equality if and only if the only nonzero $y_m$ is $m=j-{1\over2}$
and $\mu=1/(2j+1)$.  We conclude that it is impossible to have nonzero
solutions for $x_m$ and $y_m$ for more than one value of $m$.

The result is that the only solutions of the derivative
equations~(\ref{eq:derivatives}) have just one value of $m$ for which
$x_m$ and $y_m$ are nonzero, and for that value, the constraints imply
that $x_m=y_m=1$, giving $2j$ extrema of $F(x,y)$.  For all these
extrema, the two rows of $U^\dagger$ are orthogonal, as required by the
constraint we neglected.  The value of $F$ at the extreme points is a
function of $m$,
\begin{equation}
F(x,y)=F_m=
\bigl(
\sqrt{(1-\mu)r_m}+\sqrt{\mu r_{-m}}\,
\bigr)^2\;.
\end{equation}
For $0\le\mu\leq 1/(2j+1)$, $F_m$ is monotonically increasing for
$m\in[-j+{1\over2},j-{1\over2}]$.  Hence the maximum occurs at
$m=j-{1\over2}$.  Pulling all this together, we have
\begin{equation}
\label{eq:maxpU}
p_\mu(U)\le{1\over2j+1}F_{j-1/2}=
\frac{1}{2j+1}\left(\sqrt\mu+\sqrt{2j(1-\mu)}\,\right)^2=p_\mu\;.
\end{equation}
This establishes the result we needed above to complete our
determination of $\epsilon(p)$.

\subsection{Determining $\co\,\epsilon(p)$}
\label{subsec:coepsilon}

We now have to find $\co\,\epsilon(p)$.  What we show is that $\epsilon(p)$
is convex, so $\co\,\epsilon(p)=\epsilon(p)$.

\begin{figure}
\begin{center}
\includegraphics[width=8.5cm]{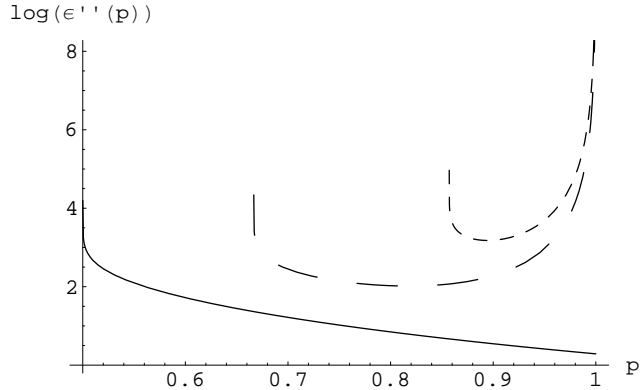}
\caption{$\log \epsilon^{\prime\prime}(p)$ for $j={1\over2}$ (solid),
$j=1$ (long-dashed), and $j=3$ (short-dashed).  We plot the log because
$\epsilon^{\prime\prime}$ diverges rapidly at $p=1$ for all $j\geq 1$.
Since the log is positive, the second derivative is bigger than $1$,
and therefore $\epsilon(p)$ is convex.
Equation~(\ref{eq:2ndderivativemin}) shows that this holds for all
$j\geq 1/2$.}
\label{fig:convexity}
\end{center}
\end{figure}

The second derivative of $\epsilon(p)$ is given by
\begin{equation}
\label{eq:2ndderivative} (p(1-p))^{3/2}\epsilon^{\prime\prime}(p)
=
\frac{\sqrt{2j}}{2j+1}\log\left(\frac{\sqrt{2jp}+\sqrt{1-p}}{\sqrt{p}-\sqrt{2j(1-p)}}\right)
- \sqrt{p(1-p)}\;.
\end{equation}
We plot $\epsilon^{\prime\prime}(p)$ for $j={1\over2}$, $1$, and $3$ in
Fig.~\ref{fig:convexity}, and we see that the second derivative is
positive for these values of $j$.  This turns out to be true for all
$j$, as we see in the following way.  It is straightforward to show
that the right-hand side of Eq.~(\ref{eq:2ndderivative}) is a
nonincreasing function in the region of interest by showing that the
derivative is bounded above by zero, and therefore its minimum value
occurs at $p=1$. We thus get the result
\begin{equation}
\label{eq:2ndderivativemin}
(p(1-p))^{3/2}\epsilon^{\prime\prime}(p)\geq
\frac{\sqrt{2j}}{2(2j+1)}\log{2j}\;,
\end{equation}
and so the second derivative of $\epsilon(p)$ is always positive for
$j\geq 1/2$, which establishes that $\epsilon(p)$ is convex.

\begin{figure}
\begin{center}
\includegraphics[width=8.5cm]{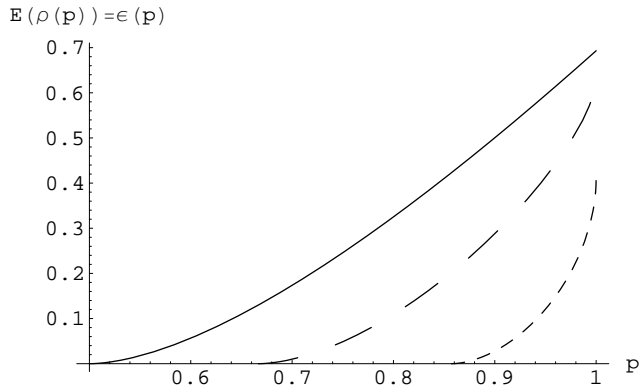}
\caption{The entanglement of formation in units of nats (we use the
natural log in the binary entropy) for $j={1\over2}$ (solid), $j=1$
(long-dashed), and $j=3$ (short-dashed). The rotationally invariant
states become less entangled as $j$ increases.}
\label{fig:eof}
\end{center}
\end{figure}

We have established that the entanglement of formation is given by
$E_F\bigl(\rho(p)\bigr)=\co\epsilon(p)=\epsilon(p)$, thus verifying
Eq.~(\ref{eq:EoF}).  We plot the $E_F\bigl(\rho(p)\bigr)$ for
$j={1\over2}$, $1$, and $3$ in Fig.~\ref{fig:eof}.

\section{Convex-Roof Measures}
\label{section:others}

Many entanglement measures can be constructed using the the
convex-roof extension.  Given any pure-state entanglement measure
$X(\psi)= f(\rho_A)$ that is (i)~invariant under local unitaries
and (ii)~a concave function of the marginal density matrix
$\rho_A$, $\co\,X(\rho)$ is an entanglement
monotone~\cite{vidal00}. The procedure used here to find the
entanglement of formation of rotationally invariant states applies
to any other convex-roof extension: first determine
\begin{equation}
\xi(p) \equiv\min_{\{|\psi\rangle\}}
\bigl(\,X(\psi)\bigm|\langle\psi|\Pi_{j-1/2}|\psi\rangle=p\,\bigr)\;,
\end{equation}
the minimum value of $X$ over all pure states that project to
$\rho(p)$, and then compute the convex hull of $\xi(p)$.  The
state $\ket\chi$ of Eq.~(\ref{eq:minstate}), which minimizes
$\epsilon(p)$, also minimizes $\xi(p)$, because different
pure-state entanglement measures give the same ordering of pure
states.  In terms of the quantities introduced in our
consideration of the entanglement of formation, this is the
statement that for the state~(\ref{eq:Schmidt}), $f(\mu)\equiv
X(\psi)$ is an increasing function of $\mu$ for
$0\le\mu\le1/(2j+1)$, thus giving $\xi(p)=f\bigl(\mu_{\rm
min}(p)\bigr)$, in analogy to Eq.~(\ref{eq:Hmumin}).

In this section we apply this technique to find the concurrence,
the tangle, and the negativity of rotationally invariant states of
a spin-$j$ particle and a spin-$1\over2$ particle.

\subsection{I-Concurrence}

The generalized concurrence~\cite{rungta01} of a joint pure state
$\ket{\psi}$ of systems $A$ and $B$ measures the purity of the
marginal states.
\begin{equation}
\label{eq:concurrence} C(\psi) =
\sqrt{2\left[1-\mbox{tr}(\rho_A^2)\right]}
=\sqrt{2\left[1-\mbox{tr}(\rho_B^2)\right]}\;.
\end{equation}
The concurrence of the state~(\ref{eq:Schmidt}) is $C(\psi)=
2\sqrt{\mu(1-\mu)}$.  The convex-roof extension of the concurrence,
$C(\rho)\equiv\co\,C(\rho)$, is called the \emph{I-concurrence}.

As discussed above, our results for the entanglement of formation
determine the function
\begin{eqnarray}
c(p)&\equiv&\min_{\{|\psi\rangle\}}
\bigl(\,C(\psi)\bigm|\langle\psi|\Pi_{j-1/2}|\psi\rangle=p\,\bigr)\nonumber \\
&=&2\sqrt{\mu_{\rm min}(1-\mu_{\rm min})} \nonumber \\
&=&\frac{2}{2j+1}\Bigl(\sqrt{2j}(2p-1)-(2j-1)\sqrt{p(1-p)}\,\Bigr)  \;.
\end{eqnarray}
The function $c(p)$ is clearly convex because it is the sum of convex
functions. When both particles have spin-$1\over2$, $c(p)$ is linear,
$c(p) = 2p-1$ for $1/2\le p\le1$, and it becomes more convex as the
spin~$j$ of the first particle increases.  Thus the I-concurrence of a
rotationally symmetric state is
\begin{equation}
\label{eq:icon} C\bigl(\rho(p)\bigr) =
\begin{cases}
0\;,&p\in[0,2j/(2j+1)],\\
c(p)\;,&p\in[2j/(2j+1),1].
\end{cases}
\end{equation}
The I-concurrence increases from 0 at $p=2j/(2j+1)$ to
$2\sqrt{2j}/(2j+1)$ at $p=1$.

\subsection{Tangle}

The tangle is the convex-roof extension of the squared
concurrence~\cite{osborne02}:
\begin{equation}
\tau(\rho)\equiv\co\,C^2(\rho)\;.
\end{equation}
Since
\begin{equation}
\min_{\{|\psi\rangle\}}
\bigl(\,C^2(\psi)\bigm|\langle\psi|\Pi_{j-1/2}|\psi\rangle=p\,\bigr)=
c^2(p)
\end{equation}
and the convexity of $c(p)$ implies the convexity of $c^2(p)$, the tangle
of rotationally symmetric states is given by
\begin{equation}
\label{eq:tangle} \tau\bigl(\rho(p)\bigr) =
\begin{cases}
0\;,&p\in[0,2j/(2j+1)],\\
c^2(p)\;,&p\in[2j/(2j+1),1].
\end{cases}
\end{equation}

\subsection{Convex-Roof-Extended Negativity}

The negativity of a joint density operator $\rho$ is proportional to
the sum of the negative eigenvalues of the partial transpose of $\rho$:
\begin{equation}
\mathcal{N}(\rho) = -2\sum_{\mu_j\leq 0}\mu_j = || \rho^{T_B}|| - 1\;.
\end{equation}
Here the $\mu_j$'s are the eigenvalues of the partial transpose
$\rho^{T_B}$ of $\rho$ with respect to system $B$, and $||\rho^{T_B}||$
is the sum of the absolute values of these eigenvalues.  The negativity
is an entanglement monotone~\cite{vidal02}, and it is straightforward
to evaluate.  The negativity does not, however, identify all entangled
states. There are states, the bound entangled states, that have a
positive partial transpose and thus zero negativity, yet are still
entangled. Notice that there are no such states in the one-parameter
family of rotationally symmetric states.  To get around the inability
of the negativity to identify all entangled states, the
convex-roof-extended negativity, $\co\,\mathcal{N}$, has been proposed
as a mixed-state entanglement measure.  This comes with a considerably
increased difficulty in computation, of the sort associated with any
convex-roof entanglement measure, but the convex-roof-extended
negativity has been evaluated for the isotropic states and Werner
states~\cite{lee03}.

For $p\in[2j/(2j+1),1]$, the partial transpose of the rotationally
symmetric state $\rho(p)$ of Eq.~(\ref{eq:invariantstate}) has two
eigenvalues
\begin{eqnarray}
\mu_+&=&\frac{1}{2(j+1)}\left(\frac{1}{2j+1}+p\right)\;,\nonumber\\
\mu_-&=&\frac{1}{2j+1}-\frac{p}{2j}\;,
\end{eqnarray}
the first of which is $2(j+1)$-fold degenerate, and the second of which
is $2j$-fold degenerate and is negative for
$p\in(2j/(2j+1),1]$~\cite{schliemann02}.  Thus the negativity of
$\rho(p)$ is
\begin{equation}
\mathcal{N}\bigl(\rho(p)\bigr) =
\max\!\left[0,2\!\left(p-\frac{2j}{2j+1}\right)\right]\;.
\end{equation}

To find the convex-roof-extended negativity, we note that for the
states~(\ref{eq:Schmidt}), $\mathcal{N}(\psi)=
2\sqrt{\mu(1-\mu)}=C(\psi)$.  This means that for rotationally
symmetric states, the convex-roof-extended negativity is actually
identical to the I-concurrence~(\ref{eq:icon}),
\begin{equation}
\co\,\mathcal{N}\bigl(\rho(p)\bigr)=
\co\,C\bigl(\rho(p)\bigr)=C\bigl(\rho(p)\bigr)\;.
\end{equation}
\section{Summary}
\label{section:summary} We have computed the entanglement of
formation of rotationally symmetric states for a $[(2j+1)\times
2]$-dimensional system. Three features of the problem simplified
the endeavor: the rotationally invariant states are specified by a
single parameter, pure states that are twirled to an invariant
state are determined by a single constraint, and only one local
unitary is needed to relate the pure states to a fiducial state.
The first makes finding the convex hull in
Eq.~(\ref{eq:cofepsilon}) easier, and the other two allow one to
efficiently characterize the subset of pure states over which one
minimizes in Eq.~(\ref{eq:epsilon}). These elements were also
vital in all other applications of the Terhal-Vollbrecht-Werner
procedure to date, i.e., generalized Werner states and isotropic
states.

The rotationally symmetric states of a spin-$j$ particle and a
spin-$1\over2$ particle are similar in some ways to the states of
a two-qubit system: positive partial transpose is necessary and
sufficient for separability, and the pure states $R\otimes
R\ket\chi$ in the optimal decomposition all have the same
entanglement. The entanglement in these states vanishes as one
spin becomes more classical ($j\rightarrow\infty$). We have also
determined the I-concurrence, tangle, and the convex-roof
negativity of the rotationally symmetric states. These all display
the same features as the entanglement of formation. This is quite
different from what occurs for the isotropic states in high
dimensions, where the concurrence is linear~\cite{rungta03}, the
tangle has the same behavior as the entanglement of
formation~\cite{rungta03}, and the convex-roof negativity is the
same as the negativity of the isotropic state~\cite{lee03}.

Since we have determined an optimal decomposition, we can also
find the entanglement of formation of states that are not
rotationally symmetric.  This procedure for extending optimal
decompositions to other states was outlined by Vollbrecht and
Werner~\cite{vollbrecht02}: arbitrary convex combinations of the
pure states $R\otimes R \ket{\chi}$ have the same entanglement as
the rotationally symmetric states, which are uniform convex
combinations of these states.  This extension procedure does not
cover the entire space; the states for which it does apply and
their properties are currently being investigated.

We have examined the simplest case of rotationally invariant
states, and the next step, which we are pursuing, is to
investigate two spin-1 particles, where the rotationally symmetric
states constitute a two-parameter family. For two spin-1
particles, it is known that the positive partial transpose
condition is necessary and sufficient for separability, and the
entanglement of a large fraction of the rotationally symmetric
states can be obtained from extending the decompositions for
generalized Werner states and isotropic states
\cite{vollbrecht02}. However the calculation for the rest of the
states poses some difficulties, because the important features
that facilitated the calculations in this paper are now absent.
The states are functions of two variables, there are two
constraints on the pure states, instead of one, and it becomes
difficult to characterize efficiently the states that are twirled
to an invariant state.  In particular, the technique used to
remove the unitary on one subsystem, thus obtaining
Eq.~(\ref{eq:fiducial}), no longer applies.  The entire problem
gets progressively more difficult as the spins of the two
particles increase.

\begin{acknowledgments}
This work was supported in part by Office of Naval Research Grant
No.~N00014-00-1-0578.
\end{acknowledgments}


\begin{thebibliography}{99}
\bibitem{schroedinger}
E.~Schroedinger, Naturwiss. {\bf 23}, 807 (1935).

\bibitem{epr}
A.~Einstein, B.~Podolsky, and N.~Rosen, Phys. Rev. {\bf 47}, 777
(1935).

\bibitem{bell}
J.~S.~Bell, Physics {\bf 1}, 195 (1964).

\bibitem{lo}
{\em Introduction to Quantum Computation and Information}, edited
by H.-K.~Lo, S.~Popescu, and T.~Spiller (World Scientific,
Singapore, 1998).

\bibitem{nielsen}
M.~A.~Nielsen and I.~L.~Chuang, {\em Quantum Computation and
Quantum Information} (Cambridge University Press, Cambridge,
England, 2000).

\bibitem{acin03}
A.~Ac\'{i}n, N.~Gisin, L.~Masanes, and V.~Scarani, Int. J. Quant.
Inf. {\bf 2}, 23 (2004).

\bibitem{bennett96-1}
C.~H. Bennett, H.~J. Bernstein, S.~Popescu, and B.~Schumacher,
Phys.\ Rev.~A\, {\bf 53}, 2046 (1996).

\bibitem{popescu97}
S. Popescu and D. Rohrlich, Phys.\ Rev.~A {\bf 56}, R3319 (1997).

\bibitem{bennett96-2}
C.~H. Bennett, D.~P. DiVincenzo, J.~A. Smolin, and W.~K. Wootters,
Phys.\ Rev.\ A {\bf 54}, 3824 (1996).

\bibitem{wootters01}
W.~K.~Wootters, Quant Inf.\ Comp. {\bf 1}, 27 (2001).

\bibitem{hayden01}
P. Hayden, M. Horodecki, and B. M. Terhal, J.\ Phys.\ A {\bf 34},
6891 (2001).

\bibitem{wootters98}
W.~K.~Wootters, Phys. Rev. Lett. {\bf 80}, 2245 (1998).

\bibitem{terhal00}
B.~M.~Terhal and K.~G.~H.~Vollbrecht, Phys.\ Rev.\ Lett.\ {\bf
85}, 2625 (2000).

\bibitem{vollbrecht02}
K.~G.~H.~Vollbrecht and R.~F.~Werner, Phys. Rev. A {\bf 64},
062307 (2002).

\bibitem{giedke03}
G.~Giedke, M.~M.~Wolf, O.~Kr\"{u}ger, R.~F.~Werner, and
J.~I.~Cirac, Phys. Rev. Lett. {\bf 91}, 107901 (2003).

\bibitem{vidal02}
G.~Vidal and R.~F.~Werner, Phys. Rev. A {\bf 65}, 032314 (2002).

\bibitem{vedral96}
V.~Vedral and M.~B.~Plenio, Phys. Rev. A {\bf 57}, 1619 (1996).

\bibitem{vidal99}
G.~Vidal and R.~Tarrach, Phys. Rev. A {\bf 59}, 141 (1999).

\bibitem{rungta01}
P.~Rungta, V.~Bu\v{z}ek, C.~M.~Caves, M.~Hillery, and G.~J.
Milburn, Phys.\ Rev.~A {\bf 64}, 042315 (2001).

\bibitem{osborne02}
T.~J.~Osborne, {\tt quant-ph/0203087}.

\bibitem{wei03}
T.-C.~Wei, P.~M.~Goldbart, Phys. Rev. A {\bf 68}, 042307 (2003).

\bibitem{virmani00}
S.~Virmani and M.~B.~Plenio, Phys. Lett. A {\bf 268}, 31 (2000).

\bibitem{schliemann02}
J.~Schliemann, Phys. Rev. A {\bf 68}, 012309 (2002).

\bibitem{bartlett03}
S.~D.~Bartlett, T.~Rudolph, R.~W.~Spekkens, Phys. Rev. A {\bf 70},
032321 (2003).

\bibitem{durkin04}
G.~A.~Durkin, C.~Simon, J.~Eisert, and D.~Bouwmeester, Phys. Rev.
A {\bf 70}, 062305 (2004).

\bibitem{hendriks}
B.~Hendriks and R.~F.~Werner [B.~Hendriks, Diploma Thesis,
University of Braunschweig, Germany, 2002] (unpublished).

\bibitem{schliemann05}
J.~Schliemann, {\tt quant-ph/0503123}.

\bibitem{breuer05a}
H-P.~Breuer, {\tt quant-ph/0503079}.

\bibitem{breuer05b}
H-P.~Breuer, {\tt quant-ph/0506224}.

\bibitem{shankar}
R.~Shankar, {\em Principles of Quantum Mechanics} (Plenum Press,
New York, 1994).

\bibitem{vidal00}
G.~Vidal, J.\ Mod.\ Opt.\ {\bf 47}, 355 (2000).

\bibitem{lee03}
S.~Lee, D.~P.~Chi, S.~D.~Oh, and J.~Kim, Phys. Rev. A {\bf 68},
062304 (2003).

\bibitem{rungta03}
P.~Rungta and C.~M.~Caves, Phys. Rev. A {\bf 67}, 012307 (2003).

\end{thebibliography}
\end{document}